\documentstyle[prb,aps,preprint,psfig]{revtex}
\begin{document}
\draft
\title{The Role of the Carrier Mass in Semiconductor
Quantum Dots}
\author{M. S. Chhabra, V. Ranjan, and Vijay A. Singh}
\address{Physics Department, I.I.T.-Kanpur, U.P. 208106, INDIA}
\date{January 1999}
\maketitle
\bibliographystyle{$HOME/bib/prsty}

\begin{abstract}
In the present work we undertake a re-examination of effective
mass theory (EMT) for a semiconductor quantum dot. We take into
account the fact that the effective mass ($m_i$) of the carrier inside
the dot of radius R is distinct from the mass 
($m_o$) in the dielectric coating surrounding the dot. The
electronic structure of the quantum dot is determined in crucial
ways by the mass discontinuity factor $\beta \equiv m_i/m_o$. In this
connection we propose a novel quantum scale, $\sigma$, which is
a dimensionless parameter proportional to $\beta^2 V_0 R^2$, 
where $V_0$ represents the barrier due to dielectric coating.
The scale $\sigma$ represents a \textit{mass} modified
``\textit{strength}'' of the \textit{potential}.  We
show both by numerical calculations and asymptotic analysis that
the charge density near the nanocrystallite surface, $\rho (r =
R)$, can be large and scales as $1/\sigma$. This fact suggests 
a significant role for the surface in an EMT based model.  We also
show that the upshift in the ground state energy is weaker than
quadratic, unlike traditional EMT based calculations, and chart
its dependence on the proposed scale $\sigma$. Finally, we 
demonstrate that calculations based on   
our model compare favorably with valence band
photoemission data and with more elaborate theoretical
calculations. 
\end{abstract}


\section{Introduction}
\label{s:intro}

	Semiconductor nanocrystallites, more popularly known as
quantum dots (QDs), have been extensively studied in the past
decade and a half. The system is interesting from the point of
view of basic physics, with the carriers being confined to an
essentially ``zero'' dimensional structure. The efficient
luminescence observed in some of these crystallites makes them 
promising candidates for opto-electronic devices. Further, the
inexorable drive towards device minituarization makes them
technologically significant.

	The physical dimensions  of a typical QD  are in the range
1-10 nm. It is coated by a dielectric which may be a polymer, glass,
oxide, etc. depending on the method of preparation. In a simplified
effective  mass theory  (EMT)   approach, the  QD is  taken  to be
spherical  with radius $R$  and the   dielectric presents a  finite
barrier   $V_0 \in     $   [1 -      10   eV]  to  the    carriers
(electrons/holes).   This  is    shown  in  Fig.~\ref{fig:qdot}.  The   carrier
confinement leads to an  enhancement of the  ``band'' gap which, in
the simple quantum  confinement model (QCM), scales  inversely with
the size  ($1/R^2$).  However, experimental observations and  more
elaborate theoretical calculations suggest that the gap dependence
is  infra-quadratic ($1/R^{\gamma}$,  $1<   \gamma  <2$).  Further,
several  workers have questioned     QCM  and suggested  that   it
downplays the significance of surface-related effects.

	In   the present  work  we   demonstrate  that  EMT can be
reconciled with the  observed infra-quadratic shift of the ``band''
edges. More importantly, an  examination of the charge  density in
our EMT based model reveals the significant role of the surface in
a direct fashion. To accomplish  this, we re-examine the EMT taking
into  account the  fact that  the   effective mass  of the carrier
inside the dot ($m_i$) is distinct from  the mass outside ($m_o$).
In other words, there is a  mass discontinuity across the potential
barrier  depicted in Fig.~\ref{fig:qdot}.  The  ratio $\beta~ (\equiv m_i/m_o)$
can  significantly affect the  electronic properties of a QD. It
is well known that the classical scale representing the surface to
volume ratio ($S/V$)  plays an important  role in determining both
classical and quantum  properties.  We  propose a novel   quantum
scale $\sigma$ which is a dimensionless parameter proportional to
$ \beta^2 V_0R^2 $ (Eq.~(\ref{eq:strength})). This scale represents a
{\it mass}  modified  ``{\it strength}''  of the
{\it potential} (MSP).   The correction to the ground state energy  $E_1$ and
the charge  density at  the  interface $\rho   (r=R)$ scale in  an
appealingly simple way with the proposed scale $\sigma $.

	In   Sec.~\ref{s:basic} we describe   the basic   model in
brief.  The carrier (electron/hole) is  confined to a spherical QD
of radius $R$  with a finite  barrier of height $V_0$ (see Fig.~\ref{fig:qdot}).
As suggested by  Brus \cite{brus83},  the BenDaniel-Duke boundary
conditions \cite{bend66} are appropriate for a  system with a mass
discontinuity across the barrier.  This aspect has been ignored in
a   number   of theoretical    calculations   in  the   past  (see
Sec.~\ref{s:disc}). As will be  shown subsequently, the physics of
the  problem is     determined  in  crucial  ways  by    the  mass
discontinuity     factor  $\beta (\equiv m_i/m_o)$. 
A pedagogical point,
we  highlight is that  the BenDaniel-Duke boundary conditions must
be  applied  to  the full   wavefunction and   not to  the partial
wavefunction. The    error   involved in  employing   the  partial
wavefunction can be substantial and is presented in Table~\ref{t:corrbound}.

	The        numerical      results  are      presented   in
Sec.~\ref{s:result}.   Previous works in   the field have devoted
considerable attention  to the upshift  of the ground state energy
$E_1$.  We focus instead on  the  charge density  $\rho (r)$. Brus
\cite{brus83} had  pointed out that   the charge density near  the
nanocrystallite surface can be large if the carrier effective mass
$m_i$   is  small.  We   show   that the   charge  density  at the
nanocrystallite  boundary $\rho (r=R)$ is  indeed large if $\beta~
(=  m_i/m_o)$ is small (Fig.~\ref{fig:chrgbeta}). A more
complete picture is provided by our proposed MSP scale 
$\sigma~(\propto \beta^2 V_0 R^2)$.
In Fig.~\ref{fig:chrgsig} we demonstrate that
$\rho (R)$ scales  inversely with  $\sigma$.
For a quantum dot, $S/V$ ratio is
large. Thus the fraction of unsaturated bonds on the surface
increases and/or the formation of molecular complexes on the
surface is favoured. The large charge density on the surface
suggests a novel non-classical reason for the importance of the
surface. We also  show  that the ground  state
energy $E_1$  exhibits a size dependence which  is weaker than the
quadratic  ($1/R^2$) dependence  one  has come to  expect from EMT
(Fig.~\ref{fig:ground}).

	A  detailed  asymptotic     analysis  is  presented     in
Sec.~\ref{s:asymp}    to    explain   the   numerical   results of
Sec.~\ref{s:result} and to present them  in terms of the proposed
quantum scale $\sigma$.  We show that $\rho (R)$ scales as $\simeq
1/\sigma $.  We introduce the   notion of the peneteration   depth
$\delta$ in this connection.  We  also explain the infra-quadratic
dependence  of the   ground  state  energy  on  the size  $R$.  In
particular, we demonstrate  that  the correction to  the quadratic
term scales as $\sim -1/\sqrt{\sigma}$.

	Section~\ref{s:disc} constitutes the discussion. We show that
our model  compares favourably  with the experimental valence-band
photoemission data of Colvin \textit{et. al.} \cite{colv91} on CdS (Fig.~\ref{fig:colvin}) and
with   more  elaborate  tight-binding  calculations  \cite{lipp89}
(Fig.~\ref{fig:lipp}).  We summarize our work and suggest directions for future
research.

\section{Basic Theory}
\label{s:basic}

We consider a spherical semiconductor nanocrystallite of diameter 
$d=2R$ (See Fig.~\ref{fig:qdot}). An additional electron in the ``conduction band'' 
of such a crystallite is described in effective mass theory by the 
Hamiltonian:
\begin{eqnarray}
H & = & -{\frac{{\hbar}^{2}}{2}}{\vec \nabla}\cdot\left({{1}\over{m^{*}
(\vec r)}}\cdot {\vec \nabla} \right)+V(\vec r)  \label{eq:ham}
\end{eqnarray}
For a position dependent mass the appropriate hermitian kinetic
energy operator is given by the first term on the right hand
side of Eq.~(\ref{eq:ham}) \cite{bend66}. The
electron effective mass inside the nanocrystallite ($m_i$) is different
from the effective mass in the dielectric coating ($m_o$). A
useful parameter is the ratio ($\beta$) of the effective masses.
\begin{eqnarray}
\beta & = & \frac{m_i}{m_o}         \label{eq:beta}
\end{eqnarray}
For the experimentally relevant cases explored in the 
last two decades, $\beta$ is less than unity. For extreme cases, such 
as GaAs or InAs quantum dots, $\beta$ equals $0.067$ and $0.02$, respectively,
where we have taken $m_{o}=m_{e}$, the free electron mass. An important 
aspect of the present study is to highlight the significance of this 
parameter $\beta$. The potential V(r) in Eq.~(\ref{eq:ham}) is
given by
\begin{eqnarray}
V(r) & = & \left\{\begin{array}{cc}
		0 &\mbox{\hspace{1 cm} $ r  \leq 0$}\\
		V_{o} & \mbox{\hspace{1 cm} $ r > R$} 
		\end{array} \right.       \label{eq:pot}
\end{eqnarray}
Here $V_{o}$ is a large positive potential and represents the dielectric 
coating surrounding the nanocrystallite (See Fig.~\ref{fig:qdot}). Typically
a nanocrystallite is surrounded by dielectrics such as glasses,
polymers, organic solvents or oxides and hydrides \cite{wang91}.
Thus, the electron 
is in a spherically symmetric well. The wavefunction for this spherically 
symmetric problem can be written in the form:
\begin{equation}
\psi_{nlm}(r)=R_{nl}(r)Y_{lm}(\theta,\phi)
\end{equation}
where the symbols have their usual meanings. It is customary to write the 
radial wavefunction $R_{nl}$ as:
\begin{equation}
u_{nl}(r)=rR_{nl} \label{eq:auxillary}
\end{equation}
The equation for the partial wavefunction $u_{nl}(r)$ is :
\begin{equation}
{{d^{2}u_{nl}}\over{dr^{2}}}-{{l(l+1)}\over{r^{2}}}u_{nl} + {{2m_{i}E}
\over{\hbar^{2}}}u_{nl}=0 \label{eq:eq_u}
\end{equation}
where $m_{i}$ is the mass of the electron inside the potential well.
The electron effective mass $m^{*}$ assumes different values $m_i$ inside 
and $m_o$ outside the well.  

For the $l = 0$ case, the standard form for the radial
wavefunction is obtained on solving Eq.~(\ref{eq:eq_u}). This is :
\begin{equation}
R_{n0}(r)= A{{\sin(k_n^{in} r)}\over{k_n^{in} r}} 
\end{equation}
for $0$ $<$ $r$ $<$ $R$ and
\begin{equation}
R_{n0}(r)=B{{e^{-k_n^{out} r}}\over{k_n^{out} r}}
\end{equation}
for $r$ $>$ $R$, where 
\begin{equation}
k_n^{in}=\sqrt{{{2 m_{i} E_{n}}\over{\hbar^{2}}}} \label{eq:kin}
\end{equation}
and,
\begin{equation}
k_n^{out}=\sqrt{{{2 m_{o} (V_{o}-E_{n})}\over{\hbar^{2}}}}
\end{equation}
and $A$ and $B$ are normalization constants.

Standard text books on quantum mechanics state the condition 
of continuity of the derivative of the wavefunction inside and 
outside the well as :
\begin{equation}
\left.{{dR_{nl}}\over{dr}} \right|_{r\rightarrow R^{-}}=\left.
{{dR_{nl}}\over{dr}} \right|_{r\rightarrow R^{+}}
\end{equation}
However, as has been discussed in the semiconductor 
literature, this condition must be replaced by the BenDaniel-Duke 
condition in case the effective masses are different across the interface. 
The condition now reads
\begin{equation}
\left.{{1}\over{m_{i}}}{{dR_{nl}}\over{dr}} \right|_{r\rightarrow R^{-}}
=\left.{{1}\over{m_{o}}}{{dR_{nl}}\over{dr}} \right|_{r \rightarrow R^{+}}
\label{eq:corrbound}
\end{equation}

A further point needs to be made at this stage. Many textbooks, 
in the process of the solution of the Schr\"{o}dinger equation,
define the 
partial wavefunction $u_{nl}$ (Eq.~(\ref{eq:auxillary})) 
and then further state that the continuity conditions at the 
interface may be imposed on $u_{nl}(r)$ \cite{schi68,powe71,schw92}. They claim that this would yield 
the same eigenvalue conditions as imposing these conditions 
on $R_{nl}(r)$ would. Quoting Schiff \cite{schi68}, `` The energy levels are 
obtained by ...(this is equivalent to making $1/R~dR/dr$ continuous there)''. 
It is shown that this is manifestly incorrect when we use the modified 
boundary condition. It is further shown that in the limit the effective 
masses become equal across the interface, the two treatments indeed become 
equivalent for this  special case. Imposition of the modified boundary condition on the full wave 
function leads, after some algebraic manipulations, to 
\begin{equation}
 k_n^{in} \cot(k_n^{in} R) = -\beta  k_n^{out} + {{1-\beta}\over{R}} \label{eq:correct}
 \end{equation}
 whereas, the use of the same condition on the partial wavefunction $u_{nl}(r)$ 
 leads to
 \begin{equation}
 k_n^{in} \cot(k_n^{in} R) = -\beta  k_n^{out}  \label{eq:wrong}
 \end{equation}
 The two conditions are evidently {\em{quite}} different. 
 However, they agree when $\beta$ is unity, recovering thereby the elementary 
 case. There are three parameters, $\beta$, $V_0$ and $R$, in the above transcendental equation. 
 As can be 
 readily seen from the two equations, the eigenvalues calculated using Eq.~(\ref{eq:correct}) 
 would deviate more and more from those calculated using Eq.~(\ref{eq:wrong}) for smaller $\beta$ and smaller $R$. The energy 
 eigenvalues obtained for a particular set of parameters is presented in 
 Table~\ref{t:corrbound}. For $\beta$=0.1, $V_0$=1.0 and $R$=20.0 $A^{o}$, the disagreement 
 is even worse, being as large as 57\%.
 In the sections that follow, we shall employ
 Eq.~(\ref{eq:correct}).

 \section{Results}
 \label{s:result}

 In this section, we present the results of our numerical calculations.
 Further, this section and the next contains the interpretation of our 
 results.  In what follows, we focus our
 attention on the spherically symmetric ($l = 0$) case.

 A quantity of central interest to us is the radial charge
 density $\rho (r)$
 \begin{eqnarray}
 \rho (r) & = & r^2 R_{n0}^2 (r)       \nonumber
 \end{eqnarray} 
 Where the multiplicative constant, $e$, the carrier charge, has
 been ignored.
 We examine the dependence of the equilibrium charge density on
 the ratio of the effective masses $\beta$ ($\equiv m_i/m_o$), the
 barrier height $V_0$ and the radius of the nanocrystallite R.
 In Fig.~\ref{fig:chrgdist}, we have plotted the charge density for different values 
 of $\beta$ and $V_0$ = 1.5 eV. The radius of the crystallite is R = 50 $A^o$.
 As can be readily seen, the value of the equilibrium charge 
 density at the crystallite boundary, $\rho(r = R)$, falls 
 rapidly with increasing $\beta$. The peak in the charge density
 occurs at $r \in [R/2, R]$; in other words, close to the surface
 of the crystallite. For small values of $\beta$, the peak is at
 $r \simeq R$. Examples of small $\beta$ are InP, InAs, GaAs,
 etc. Even if the carrier is photogenerated in the interior of
 the crystallite, it rapidly redistributes itself such that the
 equilibrium charge density is large near the boundary. In
 nanocrystallites with large $\beta (\simeq 1)$ the peak in the
 charge density shifts towards R/2. In other words r = R/2 is a 
 ``fixed point'' for the charge density.
 Figure~\ref{fig:chrgbeta} shows this more systematically. We have plotted the charge 
 density at the boundary as a function of $\beta$, and taking $V_0$ = 5 eV. There is a clear decrease 
 in the charge density as the value of $\beta$ increases.
 In the next section, we shall
  show this more rigorously in the limit of large $V_0$.
 An appropriate perspective on $\rho (r = R)$ is obtained by
 defining a \textit{mass} modified ``\textit{strength}'' of the
 \textit{potential} (MSP), $\sigma$. 
 \begin{eqnarray}
 \sigma & = & (\beta \kappa_0 R)^2          \label{eq:strength} \\
 \kappa_0 & = & \sqrt{\frac{2 m_o V_0}{\hbar^2}}    \nonumber 
 \end{eqnarray}
 Conventional textbooks \cite{schi68,math76} define the strength 
 of potential as
 $V_0 R^2$. As will become apparent in the next section, a
 more appropriate definition for the case under study is $\sigma
 = (\beta \kappa_0 R)^2 $.
 Figure~\ref{fig:chrgsig} depicts that $\rho(r)$ falls with $\sigma$. 
 This decrease is observed to be nearly linear. We shall have occasion to
 examine this universal parameter parameter $\sigma$ in the next
 section.
 In Fig.~\ref{fig:ground} we study the dependence of the ground state energy
 $E_1$ ($n = 1$, $l = 0$) of the carrier on the crystallite size
 R. We have taken $V_0$ = 5 eV for this figure.
 We observe that this dependence is infraquadratic, i.e. weaker
 than $1/R^2$. We note that this infraquadratic behaviour has
 been reported both experimentally and on the basis of
 tight-binding (TB) calculations \cite{dele93a,ranj98a}. In the
 present case infraquadratic behaviour can be traced to the
 finite nature of the barrier \textit{and} the change of the
 effective mass across the barrier. Inspection of
 Fig.~\ref{fig:ground}
 reveals that with decreasing value of $\beta$, the behaviour
 increasingly departs from the quadratic case. This is
 demonstrated rigorously in the next section on asymptotic
 analysis (Eq.~(\ref{eq:asymptotic})), where we see that 
 \begin{equation}
 E_1={{c_{1}}\over{R^{2}}} - {{c_{2}}\over{R^{3}}}
 \label{eq:ground}
 \end{equation}
 $c_{1}$ and $c_{2}$ being constants. 
 An attempt to coerce the above expression into the form
 \begin{eqnarray}
 E_1 & \simeq & \frac{C}{R^{\gamma}}
 \end{eqnarray}
 will yield $ 1 < \gamma < 2$.
 We have presented the salient results of our model in this section.
 Additional results related to experiments and earlier
 theoretical calculation on specific semiconductor
 nanocrystallites will be presented in the last section
 (Sec.~\ref{s:disc}). The following section on asymptotic analysis will
 attempt to explain the numerical results obtained in this
 section.

 \section{Asymptotic Analysis}
 \label{s:asymp}

 We now present an asymptotic analysis of Eq.~(\ref{eq:correct}) in an 
 attempt to explain the results of our calculations in the previous section.
 For an infinite potential 
 well, the eigenvalue condition reduces to :
 \begin{equation}
 k_n^{in}R=n \pi 
 \end{equation}
 where $k_n^{in}$ is defined in Eq.~(\ref{eq:kin}). We consider the case of 
 the ground state {\it{i.e.}} $n=1$. For a well which is ``sufficiently'' deep 
 but finite, we approximate the above expression by :
 \begin{equation}
 k_1^{in}R=\pi-\epsilon \label{eq:pertinf}
 \end{equation}
 where $\epsilon$ is a small number. Using Eq.~(\ref{eq:pertinf}) in
 Eq.~(\ref{eq:correct}), we obtain :
 \begin{eqnarray}
 \frac{\epsilon}{\pi} & \simeq & {{1}\over{\beta(1+\kappa_{o}R)}}
					       \label{eq:epsilon}\\
 \mbox{where \hspace{1cm}} \kappa_{o} & = & \sqrt{{{2m_{o}V_{o}}
 \over{\hbar^{2}}}}           \label{eq:kappa}
 \end{eqnarray}
 where we have assumed that $V_{o}\gg E_1$, the ground state energy.
 The condition on the smallness of
 $\epsilon$ is now apparent. If either of these parameters
 {$\beta$, R, V$_o$} is large then $\epsilon$ would be small.
 It is clear that $\epsilon$ is inversely dependent on the
 \textit{mass} modified 
 ``\textit{strength}'' of the \textit{potential} defined earlier in
 Eq.~({\ref{eq:strength})
 \begin{eqnarray}
 \epsilon & \simeq & \frac{\pi}{\sqrt{\sigma}}   \label{eq:epsig} 
 \end{eqnarray}
 when $\sqrt{\sigma} \gg \beta$.
 We are now in a position to obtain an approximate expression for
 the ground state energy $E_1$. Using Eqs.
 (\ref{eq:pertinf}) and (\ref{eq:epsilon}) and the definition of
 $k_1^{in}$ from Eq.~(\ref{eq:kin}), one obtains,
 \begin{eqnarray}
 E_{1} & = & {{\pi^{2}\hbar^{2}}\over{2m_{i}R^{2}}}\left[1-{{2}\over{\beta 
 \kappa_{o}R}}+ \ldots \right]  \label{eq:asymptotic} \\
  & = & \frac{\pi^2 \hbar^2}{2 m_i R^2} \left[1 -
  \frac{2}{\sqrt{\sigma}} + \ldots \right] \label{eq:groundsig}
 \end{eqnarray}
 Thus one can see that the ground state energy has the size dependence
 depicted in Fig.~\ref{fig:ground}, namely,
 \begin{eqnarray}
 E_1 & = & \frac{C_1}{R^2} - \frac{C_2}{R^3} \label{eq:gnd} \\
   & \simeq & \frac{C}{R^{\gamma}}   \label{eq:coerce}
 \end{eqnarray}
 with effective exponent $\gamma < 2$, as mentioned earlier.
 The EMT literature on nanocrystallite semiconductors commonly
 quote $\gamma = 2$. However, absorption and luminescence
 experiments as well as the tight-binding calculations 
 \cite{dele93a,ranj98a} yield
 $\gamma < 2$ . We have thus formally demonstrated 
 how even within the EMT
 approach, an infraquadratic exponent ($\gamma < 2$) is 
 obtained. The calculation depicted in Fig.~\ref{fig:ground}
 attests to this.

 Garrett \cite{garr79} has proposed a penetration depth for the one dimensional
 finite well problem with $\beta$ = 1. We define an analogous
 penetration depth ($\delta$) for the carrier in the three
 dimensional potential well and for the general case $\beta \neq
 1$.
 \begin{eqnarray}
 \delta & = & \frac{1}{\beta \kappa_{o}}  \nonumber \\
	& = & \frac{R}{\sqrt{\sigma}}  \label{eq:pendepth}
 \end{eqnarray}
 Next we shall examine the charge density at the surface. An
 appealing, intuitive way to understand the nature of $\rho(r =
 R)$ is to use the concept of the penetration depth.
 It is easy to see from Eq.~(\ref{eq:pendepth}) that the penetration depth will 
 be large in case $\beta$ is small. This is already a confirmation of the 
 result that the charge density is large at the boundary for small
 $\beta$ (Figs.~\ref{fig:chrgdist} and \ref{fig:chrgbeta}). Thus one
 would expect that when the carrier mass in the semiconductor is
 small (InSb, GaAs, CdS) then the charge density at the interface
 will be large.
 The penetration depth $\delta$ indicates the extent of the wavefunction
 penetration into the forbidden region. 

 We can present a more rigorous analysis of $\rho(r = R)$ by examining 
 the wavefunction. The normalized 
 wavefunction inside the potential well is :
 \begin{eqnarray}
 R_{10}(r) & = & A{{\sin (k_1^{in}r)}\over{k_1^{in}r}} \label{eq:wfn}
 \\
 \mbox{where \hspace{1cm}} A & = & {{k_1^{in}}\over{\sqrt{4\pi}}}\left[{{R}\over{2}}\left(1-{{\sin(2 
 k_1^{in}R)}\over{2 k_1^{in} R}}\right)+{{1-\cos(2 k_1^{in} R)}\over{4 k_1^{out}}}
 \right]^{-\frac{1}{2}}
 \end{eqnarray}
 Using Eq.~(\ref{eq:pertinf}) in Eq.~(\ref{eq:wfn}), we obtain after some 
 algebraic manipulations:
 \begin{eqnarray}
  R_{10}(r) & \approx & A \left(1 + \frac{\epsilon}{\pi} \right)
  \left[\frac{\sin(\pi r/R)}{\pi r/R} - \frac{\epsilon}{\pi}\cos
  \left(\frac{\pi r}{R}\right)\right]
  \label{eq:pertwfn} 
 \end{eqnarray}
 whence we get for the charge density $\rho$,
 \begin{equation}
 \rho(\epsilon,r)  \equiv  r^{2} R_{10}^{2}(r)  =  A^{2}\left(1+{{\epsilon}
 \over{\pi}}\right)^{2}\left[\left({{R}\over{\pi}}\right)^{2}\sin^{2}
 \left({{\pi r}\over{R}}\right)+ \left({{r \epsilon}\over{\pi}}\right)^{2}
 \cos^{2}\left({{\pi r}\over{R}}\right)-\left({{\epsilon r R}\over{\pi^{2}}}
 \right)\sin\left({{2\pi r}\over{R}}\right)\right] \label{eq:pertchrg}
 \end{equation}
 The charge density at the interface is:
 \begin{equation}
 \rho(R) = \left.r^{2}R_{10}^{2}(r)\right|_{r=R}=A^{2}\left(1+{{\epsilon}\over{\pi}}
 \right)\left({{\epsilon}\over{\pi}}\right)^{2} R^2 
 \label{eq:boundchg}
 \end{equation}
 Recall that $\epsilon^2 \sim 1/\sigma$ ($\equiv 1/(\beta \kappa_o
 R)^2)$. It is therefore clear that $\rho(r = R)$ would fall with
 increasing $\beta$. This prediction is in agreement with 
 Fig.~\ref{fig:chrgbeta} where $\rho(R)$ is 1/$\beta^{1.85}$.

 Within the range of validity of approximation in
 Eq.~(\ref{eq:pertinf}), $\epsilon$ is much less than unity.
 Therefore, Eq.~(\ref{eq:boundchg}) may be written as 
 \begin{eqnarray}
 \rho (R) & \approx & A^2 \frac{R^2}{\sigma} \label{eq:rhosig}
 \end{eqnarray}
 One can thus understand the inverse dependence of $\rho(R)$ on $\sigma$ in
 Fig.~\ref{fig:chrgsig}. It is interesting to note that the charge density can
 be related to the penetration depth defined in
 Eq.~(\ref{eq:pendepth}),
 \begin{eqnarray}
 \rho(R) & \approx & A^2 \delta^2       \nonumber
 \end{eqnarray}
 One may also obtain the peak position, $r_{peak}$, in the charge 
 density by a simple differentiation of $r R_{10} (r)$, where
 $R_{10} (r)$ is given by Eq.~(\ref{eq:pertwfn}).
 This yields
 \begin{equation}
 \left({{\pi r_{peak}}\over{R}}\right)\tan\left({{\pi r_{peak}}\over{R}}
 \right) = -\frac{1-\epsilon/\pi}{\epsilon/\pi} \label{eq:peakchrg}
 \end{equation}
 It can be seen easily, either by plotting Eq.~(\ref{eq:pertchrg}) or by 
 solving Eq.~(\ref{eq:peakchrg}) numerically, that the peak in the charge 
 density $\rho$ is located at values of $ R/2 \leq r \leq R$ and shifts to  
 $r=R/2$ as the value of $\epsilon$ is decreased. Decrease in 
 $\epsilon$ is brought about by an increase in  
 $\sigma$.
 Equation~(\ref{eq:peakchrg}) indicates that $r_{peak} \in [R/2, R]$.
 When $\beta$ is small, $r_{peak} > R/2$ and this is also
 graphically
 borne out by Fig.~\ref{fig:chrgdist}. Since the peak is
 shifted towards the crystallite boundary, the charge density at
 the boundary becomes substantial. When the MSP scale $\sigma$
 is large, Eq.~(\ref{eq:peakchrg}) implies that $r_{peak}
 \rightarrow R/2$. In other words there is a
 \underline{fixed-point} or ``attractor'' at r = R/2. The peak in the
 wavefunction does not shift to a value below R/2, no matter how
 large $\sigma$ is. 

 \section{Discussion}
 \label{s:disc}

 The objective of the present work is to explore EMT in the
 context of the variation of the carrier effective mass across
 semiconductor - dielectric interface in a QD. We consider a simple
 Hamiltonian described by Eq.~(\ref{eq:ham}) in order to illustrate
 the essential physics. It would be worthwhile to compare our results
 with other theoretical calculations and experiments. Below we present
 two such comparisons.

 We compare our results with the valence band photoemission
 experiment carried out by Colvin \textit{et. al.} on CdS quantum
 dots \cite{colv91}. Their work represents the first non-optical observation of
 the electronic structure of semiconductor nanostructures. The
 experiment measures the energy shift of the valence band
 maximum as the cluster size decreases, taking the largest cluster of R
 = 35 $A^o$ as a reference. For this case \cite{lipp89}, $\beta$ = 0.53 and
 Fig.~\ref{fig:colvin} depicts the dependence of the valence band shift on the
 cluster size as calculated by us. We fit our data to
 C/$R^{\alpha}$. We get an exponent $\alpha = 1.26$, which is in
 good agreement with the experimental observation.
 Colvin \textit{et. al.} attribute this
 shift to two factors : (i) the kinetic energy enhancement due to
 the quantum confinement of the hole ; (ii) the polarization
 of the crystallite and the loss of dielectric solvation energy
 \cite{brus83,brus84}. As is clear from Fig.~\ref{fig:colvin} a finite barrier quantum
 confinement model with appropriate effective masses provides a
 reasonable explanation for the experimental behaviour. We also
 note in passing that Takagahara has argued that the polarization
 term can in general be neglected \cite{taka93}.

 Lippens and Lannoo \cite{lipp89} have theoretically studied the
 same system. Our model can be compared with their results
 for the valence band shift. Fig.~\ref{fig:lipp} depicts such a
 comparison. We have taken the barrier height $V_0$ = 2 eV.
 A majority of the literature claim that TB calculations are in
 better agreement with experiments. The EMT with quadratic
 dependence on size ($\sim 1/R^2$) represents an extreme case. As
 can be seen from Fig.~\ref{fig:lipp} our EMT based model is in close
 agreement with the experimental observations. On the other
 hand,
 the standard EMT calculation of Lippens and Lannoo is at
 variance with their TB calculation.

 The majority of the theoretical work for the absorption and
 luminescence in quantum dots takes excitonic effects into account.
 In other words, the Hamiltonian involves both a hole and an
 electron. We propose to extend our simple EMT model to the
 excitonic case as well as to study excited states in the
 future. We also plan to study the capacitance of QDs by
 analyzing multi-electron effects in our model.

 We emphasize that the EMT calculation should be carried out with
 the correct boundary condition as presented in
 Eq.~(\ref{eq:corrbound}). This fact has been pointed out by
 Brus \cite{brus83}. The use of this boundary condition leads to
 an eigenvalue condition (Eq.~(\ref{eq:correct})) which is
 different from the one normally encountered in textbooks
 \cite{schi68,powe71,schw92}
 (Eq.~(\ref{eq:wrong}) with $\beta$ = 1). A further point to note is that the 
 boundary condition must be imposed on full wavefunction
 $R_{nl}(r)$. Imposing it on the partial wavefunction $u_{nl}(r)$
 would lead to incorrect results.

	EMT based calculations have normally been carried out using
an infinite barrier \cite{wogg97}. In this case, the charge density
at  the  nanocrystallite boundary  is  zero.   It  is  therefore not  
surprising  that   the
significance of the  surface is downplayed.   Further, workers who
recognize the finite  nature of the  barrier have ignored the mass
discontinuity   across  the barrier
\cite{kaya90,nomu91,shim97,lahe97}  and  the reason  sometimes  stated is that  most  of  the
population density of the carrier is still confined well within QD
\cite{kaya90}.  The   LDA  based   calculations  on  dot
capacitance and shell    filling  effects also ignore  the    mass
discontinuity  \cite{macu97}. Porous silicon,  which is  a disordered
agglomeration of   silicon nanocrystallites, presents  a  
case-study  for  the  debate  between the quantum   confinement and the
surface state schools of thought. This debate has been reviewed by
a  number of  workers \cite{kane95,prok96,cull97,john97b}. In this work we have
shown that the charge density can be large  at the surface even in
an EMT   based  framework.  This  is more   so  for semiconducting
materials  with  small   carrier effective masses   and dielectric coatings
representing a  small potential barrier. Hence, the  contention that
surface related effects are significant appears valid.

Another demerit of the infinite barrier model is its prediction
of an inverse quadratic (1/$R^2$) shift of the band gap. We have
already stated that this does not agree with experiments and with
more elaborate theoretical calculations
\cite{dele93a,ranj98a,fu97,ogut97}. We
have shown through our work that a calculation for finite
barrier model leads to a good agreement with the experimental
infraquadratic dependence.

 The highlights of the work are :
 \begin{enumerate}
 \item Ground state energy within EMT scales infraquadratically
 with the crystallite size (Eq.~(\ref{eq:asymptotic}) -- 
 (\ref{eq:coerce})).
 \item The parameter of relevance is our proposed \textit{mass} 
 modified ``\textit{strength}'' of the \textit{potential}, the MSP scale
 $\sigma = (\beta \kappa_o R)^2$. Our proposed parameter
 $\sigma$
 presents a more complete picture of the physical situation than
 $\beta$ alone does \cite{brus83}.
 \item The charge density at the boundary $\rho(R)$ can be large
 (Figs~(\ref{fig:chrgdist}) -- (\ref{fig:chrgsig})) and scales inversely 
 with $\sigma$. This is
 unlike what one normally expects from EMT.
 In the literature of semiconductor nanocrystallites there has been
 a longstanding debate between theories for quantum confinement versus
 surface states \cite{kane95,prok96,cull97,john97b}. The present 
 work provides a fresh perspective
 on the importance of surface related phenomena.
 \end{enumerate}

 \noindent \textbf{Acknowledgement }\\
 This work was supported by the Department of
 Atomic Energy through the Board of Research in Nuclear Sciences,
 India (No. 37/11/97-R.\&D.II). Discussions with Dr. Swapan Ghosh of
 BARC are gratefully acknowledged.



 \begin{table}
 \vskip 1cm
 \caption{Comparison of eigenvalues obtained by applying the correct boundary
 condition on the full wavefunction (Eq.~(\ref{eq:correct})) and on
 the partial wavefunction (Eq.~(\ref{eq:wrong})). The values of the parameters
 employed are $\beta = 0.1$, $V_0$ = 2.5 eV and R = 40 $A^o$.}
 \vskip 1cm
 \begin{tabular}{|c|c|c|}
 Eigenvalues with & Eigenvalues with &   \\
 full wavefunction & partial wavefunction & Percentage
 difference \\ \hline
 0.13 & 0.15 & 12.04  \\ \hline
 0.61 & 0.65 & 3.21  \\ \hline
 1.52 & 1.56 & 2.53  \\ 
 \end{tabular}
 \label{t:corrbound}
 \end{table}

\begin{figure}
\caption{The left side of the figure depicts an idealized
spherical quantum dot (QD) surrounded by a dielectric coating.
This is modeled by a potential well of height $V_0$. The carrier
effective mass is $m_i$ inside the well and $m_o$ in the
dielectric coating outside.}
\label{fig:qdot}
\end{figure}

\begin{figure}
\caption{A typical normalized charge density $\rho (r)$ inside
a crystallite of size R = 50 \AA. Note that for small $\beta = 
(m_i/m_o)$, $\rho (r)$ is large at the crystallite surface (r = R). As
$\beta$ increases, the position of the peak $r_{peak}$ in $\rho$
shifts to R/2.}  
\label{fig:chrgdist}
\end{figure}

\begin{figure}
\caption{The charge density at the crystallite
boundary, $\rho (r = R)$. The charge density $\rho (R)$ falls with increasing $\beta$.
We fit our data for $\rho (R)$
to $C_1/\beta^{\alpha}$, with $\alpha$ = 1.85. This is in good agreement
with the prediction of Eq.~(\ref{eq:boundchg}). $C_1$ is a constant.}
\label{fig:chrgbeta}
\end{figure}

\begin{figure}
\caption{The dependence of the charge density $\rho (R)$ at the
crystallite boundary on the proposed quantum scale $\sigma$
$(\equiv (\beta \kappa_0 R)^2)$. The scale $\sigma$ represents a
\textit{mass} modified ``\textit{strength}'' of the \textit{potential} (see
text). The data for $\rho(R)$ fit well to $C_2/\sigma^{\alpha}$
with $\alpha$ = 0.98, and in agreement with the prediction of 
Eq.~(\ref{eq:rhosig}). $C_2$ is a constant. The dotted line is a
fit to the calculated data (diamonds).}
\label{fig:chrgsig}
\end{figure}

\begin{figure}
\caption{The ground state energy $E_1$ versus the
crystallite size R. The calculated data are for four different values of
$\beta$ and are fitted to C/$R^{\gamma}$, where C is a constant. 
The values of $\gamma$ 
for the different values of $\beta$ = 2.0, 0.4, 0.1 and 0.01 are 
respectively 1.95, 1.76, 1.39 and 1.03.
We can see that the energy shift is increasingly
infraquadratic for smaller values of $\beta$.} 
\label{fig:ground}
\end{figure}

\begin{figure} 
\caption{The data 
taken from the valence band photoemission spectra for CdS
crystallite due to Colvin \textit{et. al.} (Ref.~3) is indicated
by diamond symbols. Our
calculation is indicated by the dotted line. A fit to
C/$R^{\alpha}$ leads to $\alpha$ = 1.26 for both cases. 
Our calculation gives a much improved
result as compared to traditional EMT with $\alpha$ = 2. See
Sec.~\ref{s:disc} for further discussion.}
\label{fig:colvin}
\end{figure}

\begin{figure}
\caption{Valence band shift with size of the CdS crystallite. Our
results (solid line) are very close to the tight-binding (TB) 
calculations (diamonds) of Lippens and Lannoo
(Ref.~4). In contrast, there is less agreement between the EMT (dashed
line) and TB (diamond symbols) calculations of Ref.~4.} 
\label{fig:lipp}
\end{figure}

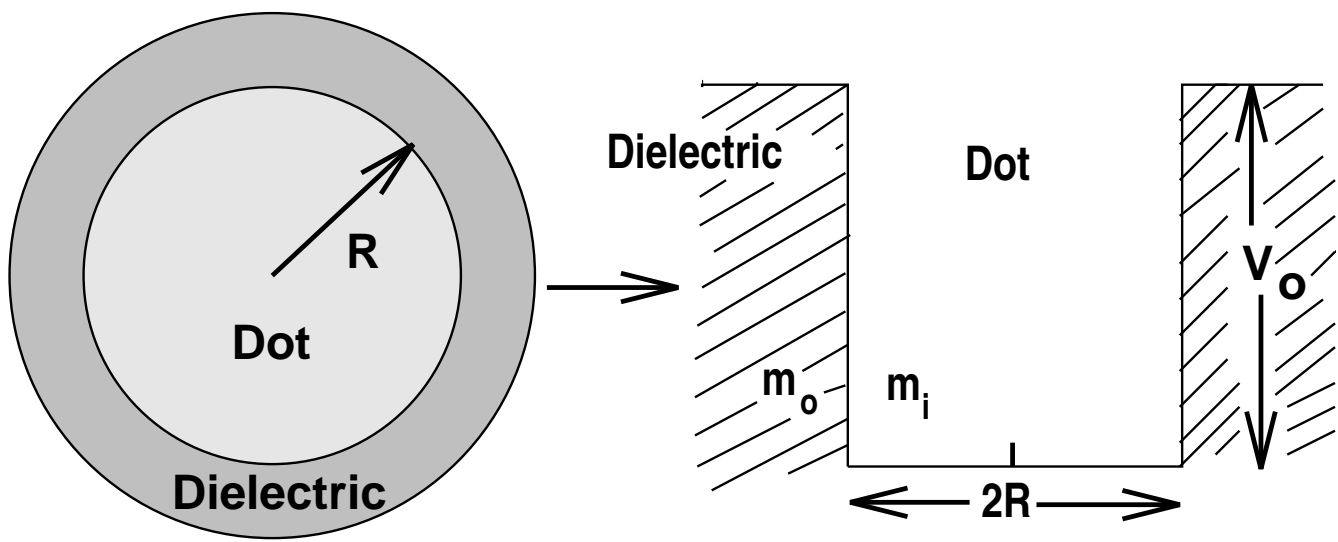
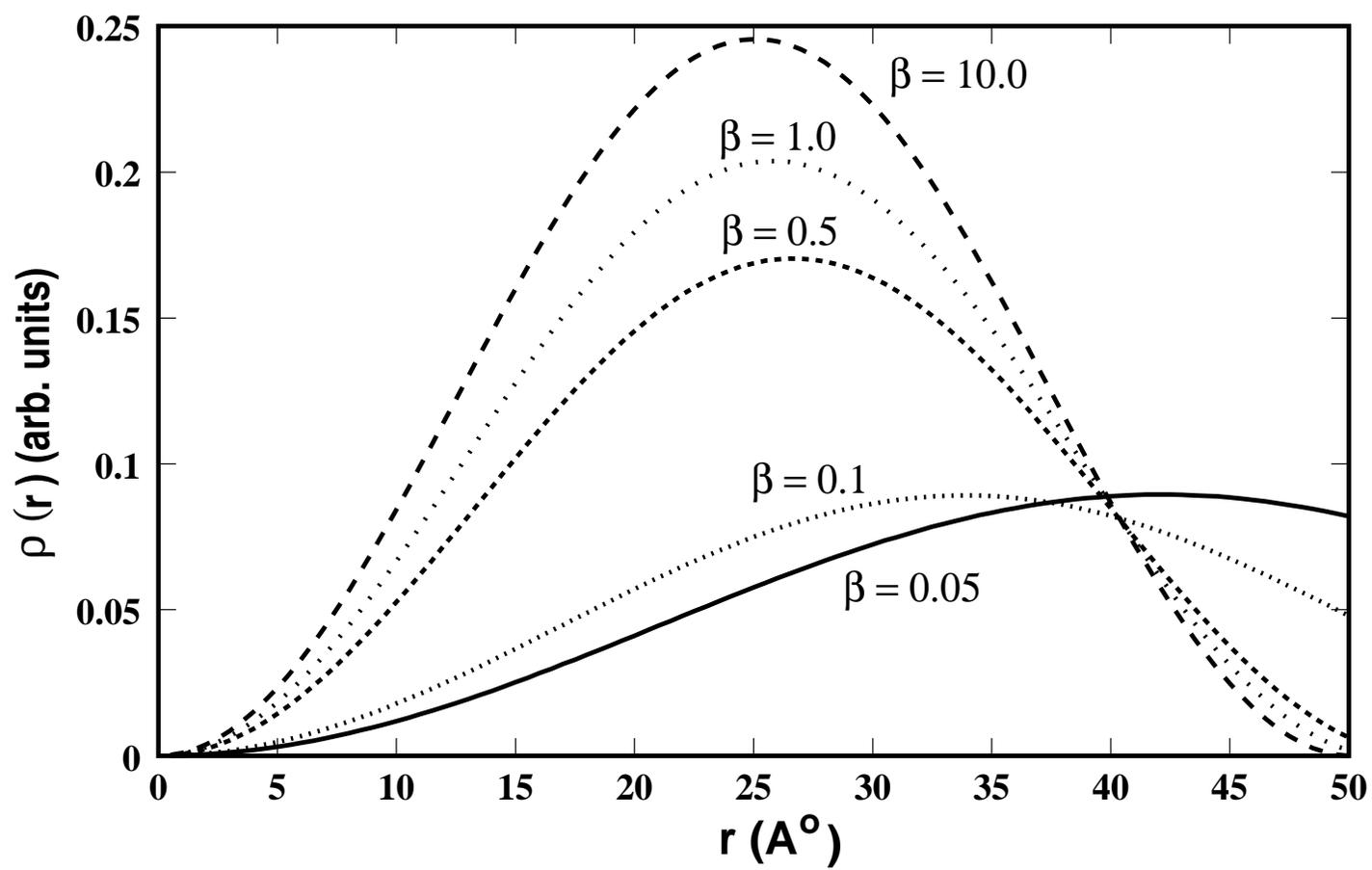
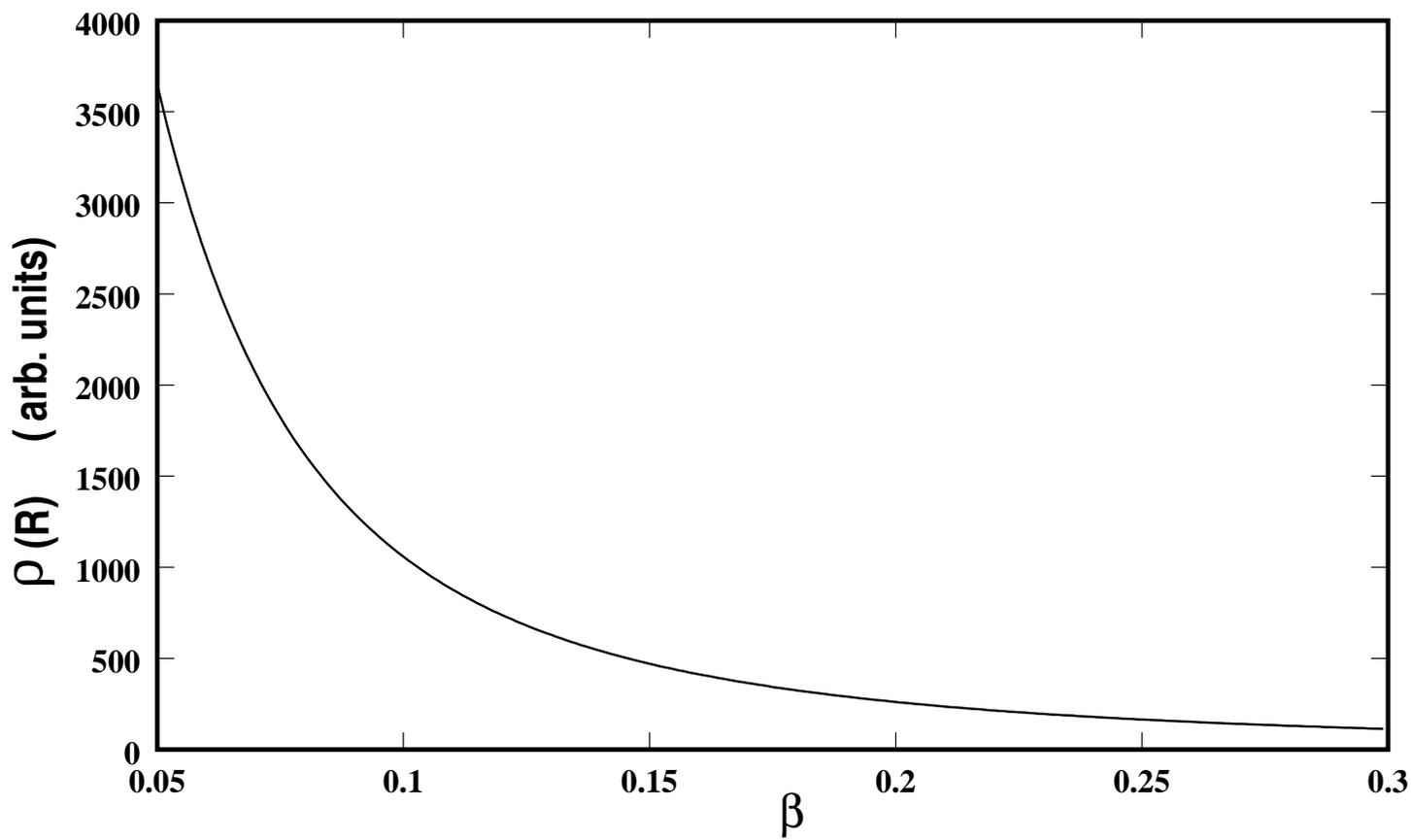
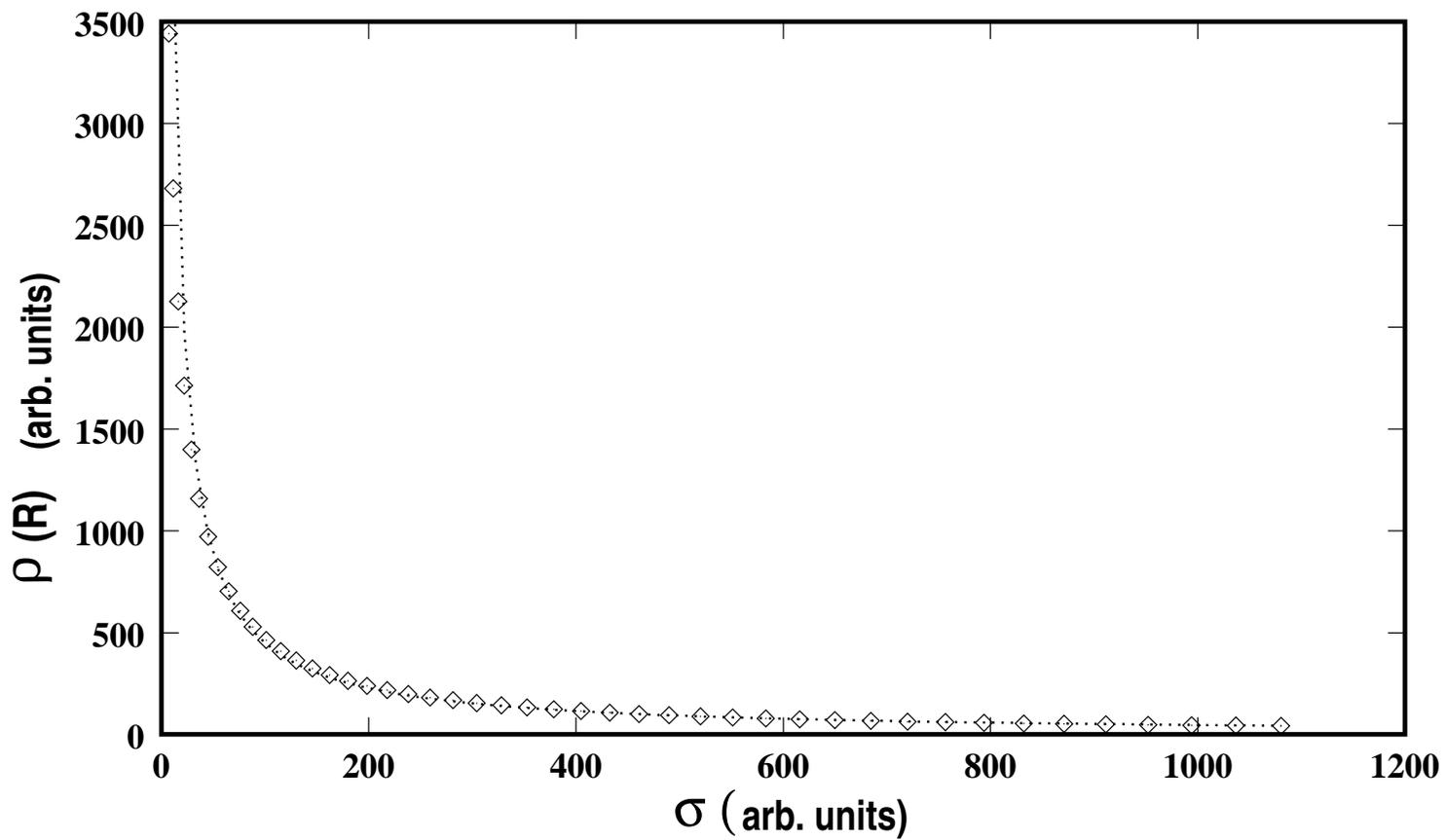
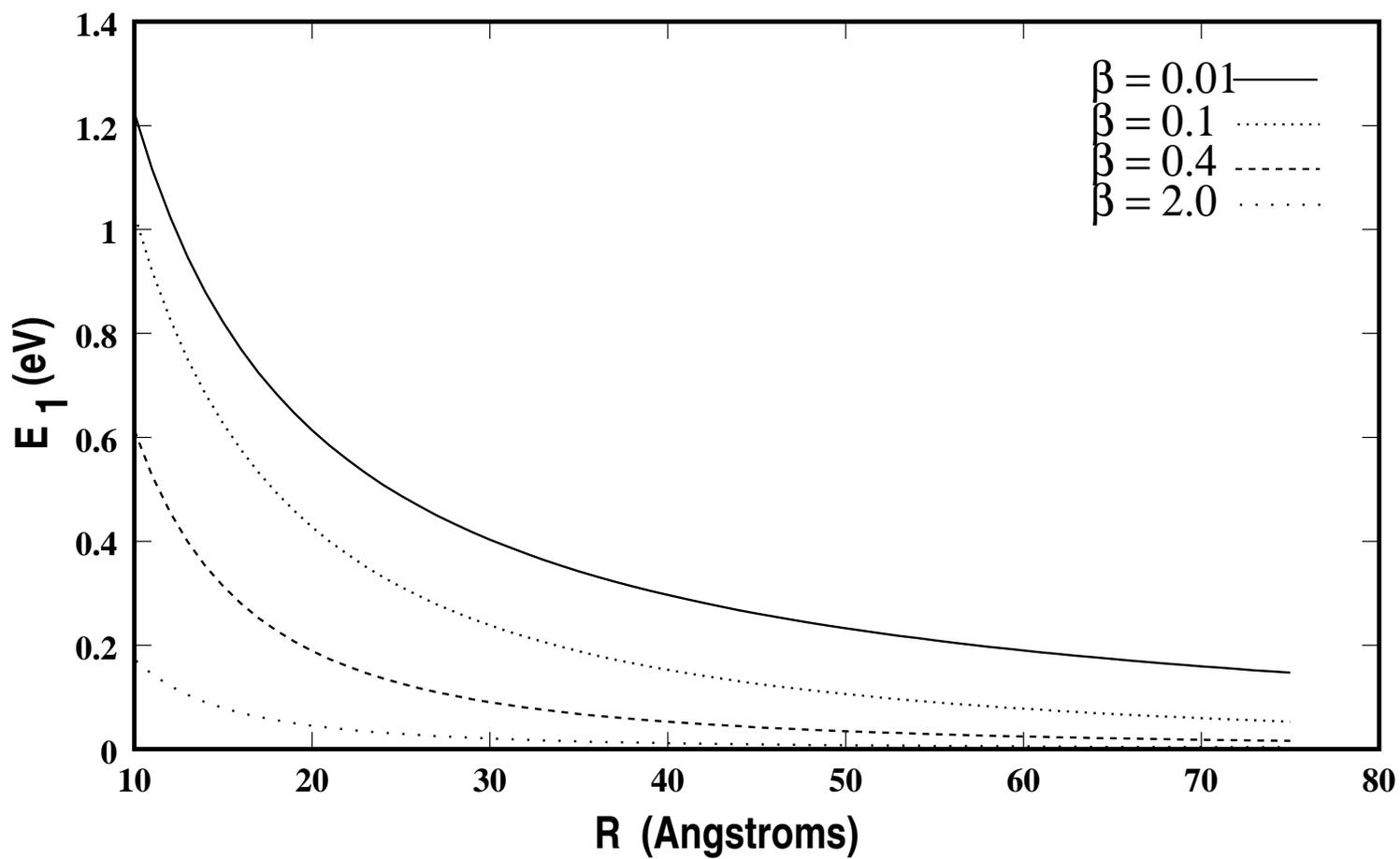
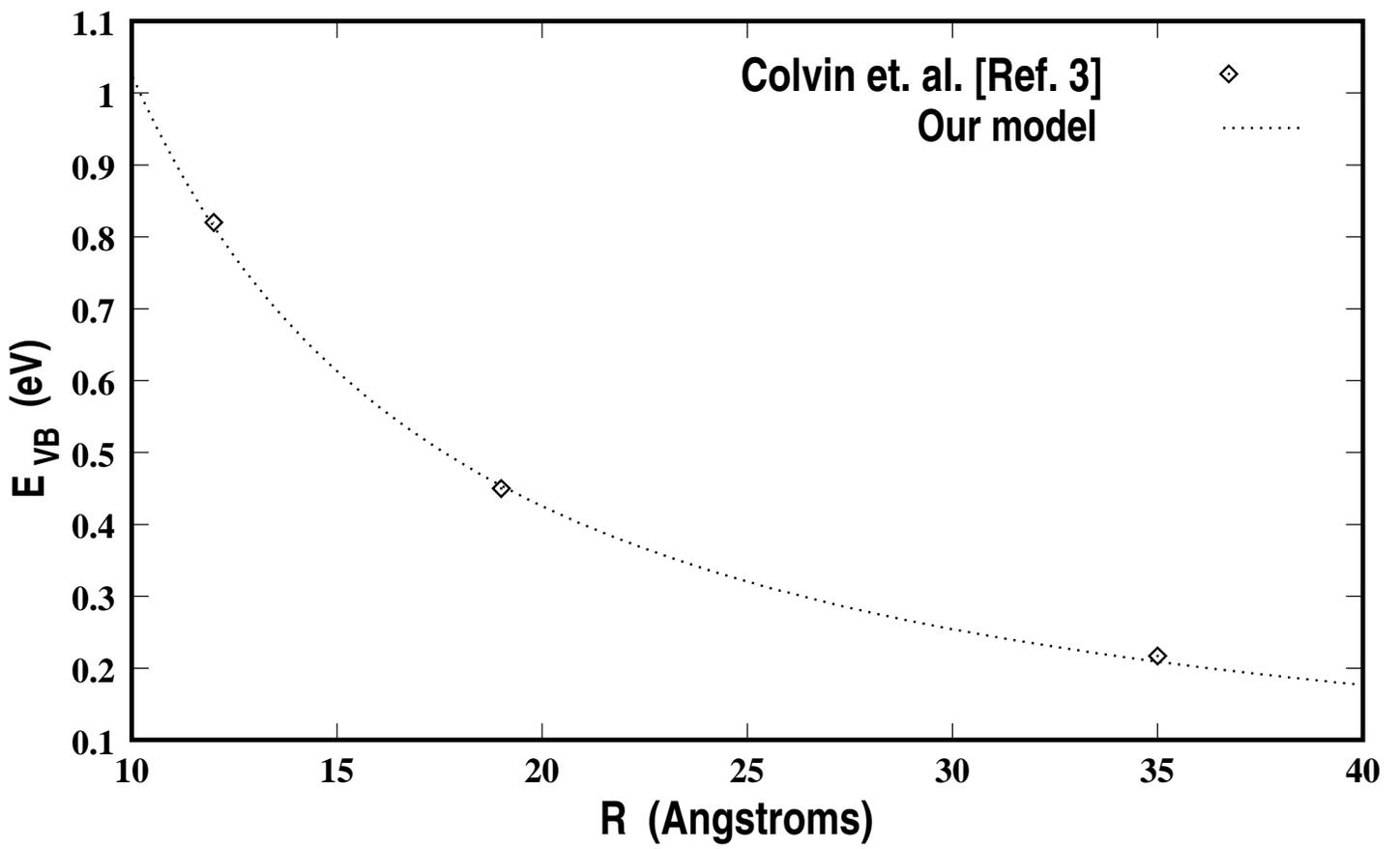
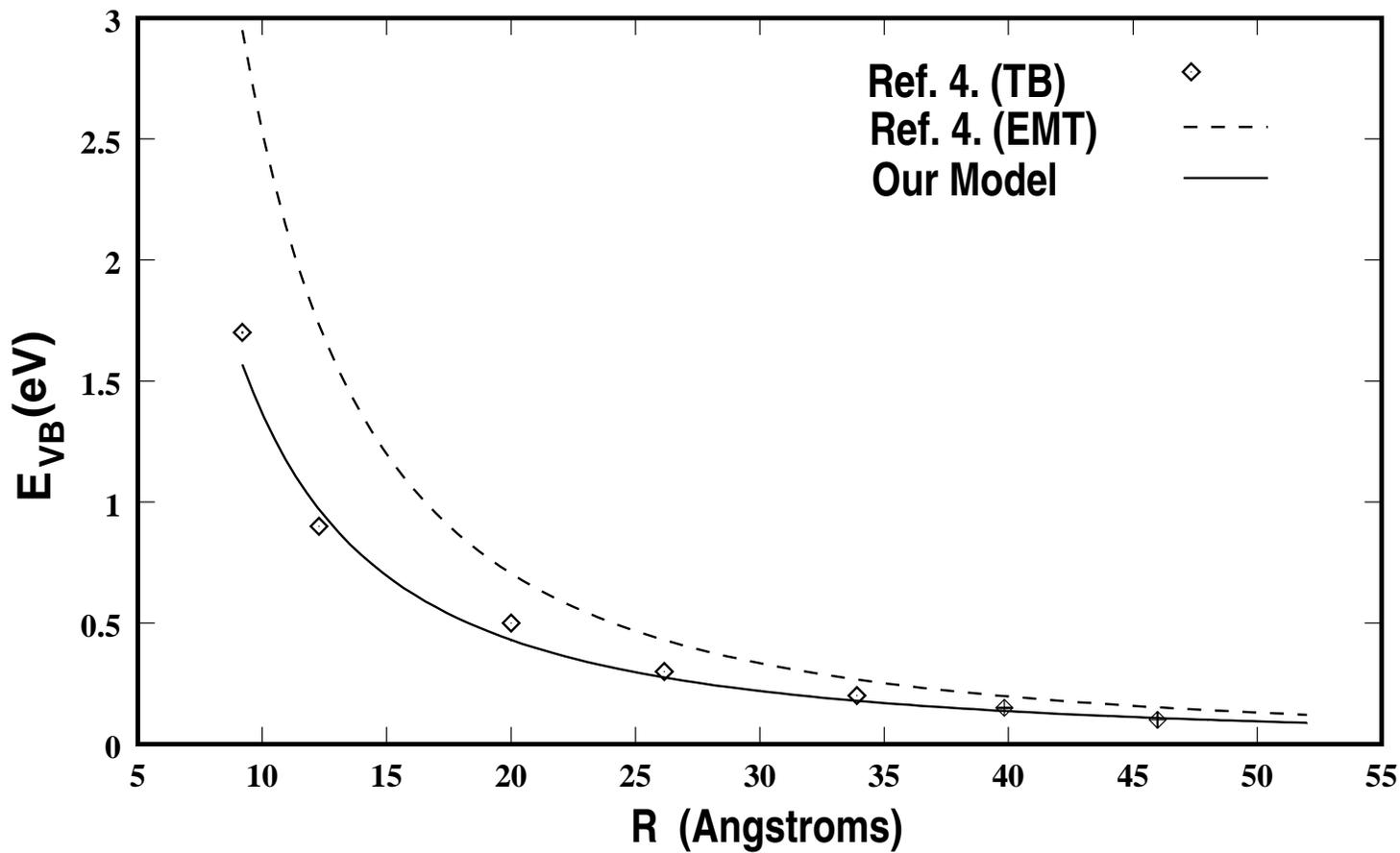
\end{document}